\documentclass{jaa}
\usepackage{natbib}
\usepackage{url}
\usepackage{graphicx}
\usepackage[multiple]{footmisc}
\usepackage{hyperref}

\newcommand{\photu}{photon units}

\newcommand{\galex}{{\it GALEX}}

\newcommand{\voyager}{{\it Voyager}}
\newcommand{\lya}{Ly$\alpha$}

\newcommand{\ebv}{E(B~-~V)}
\newcommand {\FUSE}   {{\it FUSE}} 
\newcommand{\asec}{\hbox to 1pt{}\rlap{$^{\prime\prime}$}.\hbox to 2pt{}}
\newcommand{\amin}{\hbox to 1pt{}\rlap{$^{\prime}$}.\hbox to 1pt{}}
\newcommand{\adeg}{\hbox to 1pt{}\rlap{$^{\circ}$}.\hbox to 2pt{}}
\newcommand{\degree}{\hbox to 1pt{}\rlap{$^{\circ}$}\hbox to 2pt{}}

\begin{document}

\title{Observations of High Galactic Latitude Line and Continuum Emission (912 - 1600 \AA) with New Horizons}

\author{Jayant Murthy\textsuperscript{1}, James Overduin\textsuperscript{2}, Seth Redfield\textsuperscript{3}, J. Michael Shull\textsuperscript{4,5}, 
Joel Wm. Parker\textsuperscript{6}, 
Jon P. Pineau\textsuperscript{7}, 
Anne J. Verbiscer\textsuperscript{8}, 
Richard C. Henry\textsuperscript{9}}
\affilOne{\textsuperscript{1}Indian Institute of Astrophysics, Bengaluru 560 034, India.\\}
\affilTwo{\textsuperscript{2}Towson University, Towson, Maryland 21252, USA.\\}
\affilThree{\textsuperscript{3}Astronomy Department and Van Vleck Observatory, Wesleyan University, Middletown, CT 06459, USA\\}
\affilFour{\textsuperscript{4}Department of Astrophysical \& Planetary Sciences, CASA, University of Colorado, Boulder, CO 80309, USA.\\}
\affilFive{\textsuperscript{5}Department of Physics \& Astronomy, University of North Carolina, Chapel Hill, NC 27599, USA.\\}
\affilSix{\textsuperscript{6}Department of Space Studies, Southwest Research Institute, 1301 Walnut Street, Suite 300, Boulder, CO 80302, USA.\\}
\affilSeven{\textsuperscript{7}Stellar Solutions, Aurora, CO 80011, USA.\\}
\affilEight{\textsuperscript{8}Department of Astronomy, University of Virginia, Charlottesville, VA 22904, USA.\\}
\affilNine{\textsuperscript{9}Johns Hopkins University, Dept. of Physics and Astronomy, Baltimore, MD 21218, USA.\\}


\twocolumn[{

\maketitle

\corres{jmurthy@yahoo.com}

\msinfo{1 January 2015}{1 January 2015}

\begin{abstract}
We present observations of the cosmic ultraviolet background (CUVB) from 912 – 1600 \AA\ using the Stem aperture of the Alice spectrograph on the New Horizons spacecraft at 56 AU from the Sun, providing a spectral resolution of 9 \AA\ for diffuse sources. We detect emission lines of C~III (977 \AA) and C~IV (1548/1551 \AA) at the $3 \sigma$ level with strengths of $4200 \pm 1500$ and $4100 \pm 1200$ ph cm$^{-2}$ s$^{-1}$ sr$^{-1}$, respectively, and a marginal detection of O~VI (1032/1038 \AA) at $1400 \pm 1300$ ph cm$^{-2}$ s$^{-1}$ sr$^{-1}$. We report a $3 \sigma$ detection of an emission line at 1135 Å, which we have identified with the N I resonance triplet. Although this line had earlier been observed in FUSE and SPEAR data, it had been attributed to airglow or instrumental effects. We confirm, for the first time, that it must originate in the Galaxy. 

The dust-scattered continuum is dominated by a small number ($N < 100$) of O9 – B2 stars and shows the deep absorption feature near 1000 Å seen in the stellar spectrum. Our models suggest an albedo of $a < 0.5$ over most of the spectrum (950 -- 1550 \AA) for the dust grains with the phase function asymmetry of $g < 0.6$. We find an offset, comprising the extragalactic background light and halo contributors, consistent with our earlier results from the Box, including the decline in the offset near the Lyman limit. We confirm that much of the emission must be from an unidentified component of the CUVB.

\end{abstract}

\keywords{Ultraviolet astronomy (1736), Cosmic background radiation (317), Diffuse radiation (383)}

}]


\doinum{12.3456/s78910-011-012-3}
\artcitid{\#\#\#\#}
\volnum{000}
\year{0000}
\pgrange{1--}
\setcounter{page}{1}
\lp{1}

\section{Introduction}\label{sec:intro}
We have reported on observations of the cosmic ultraviolet background (CUVB) made at high Galactic latitudes with the Alice spectrograph on the {\it New Horizons} (NH) mission \citep{Murthy2025_alice}. These were made from a distance of 56 AU from the Sun, beyond most heliospheric emission \citep{zemcov_nh2018}. We focused on the ``Box'' spectra in that work, which provided high signal-to-noise spectra but at low spatial ($2^{\circ} \times 2^{\circ}$) and spectral (172 \AA\ for diffuse sources) resolution. We found that the observed CUVB could be separated into two components: a linear component proportional to the reddening (\ebv) and a constant offset. The former is identified with dust-scattered starlight from hot stars in the Galactic Plane, while the offset is comprised of the extragalactic background light (EBL) and other line and continuum contributors \citep{Murthyreview2009}.

\begin{figure}
    \includegraphics[width=3in]{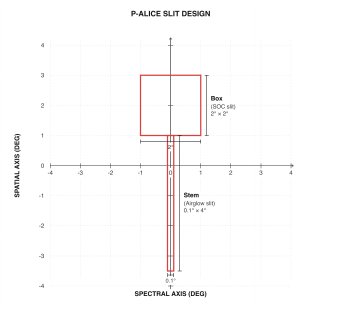}
    \caption{The Alice entrance aperture is a square Box on top of a narrow, rectangular Stem \citep{Stern2008}.}
    \label{fig:alice_slit}
\end{figure}

Simultaneous observations were made through the ``Stem'', integrating over a $4^{\circ} \times 0\adeg 1$ patch of the sky adjacent to, but not overlapping, that observed by the Box (Fig. \ref{fig:alice_slit}). The smaller Stem slit offers a much higher spectral resolution (9 \AA) for aperture-filling, diffuse sources but with a solid angle only 5\%\ that of the Box. Despite the poorer signal-to-noise, we have used Stem spectra to investigate the spectrum of the CUVB from 912 -- 1600 \AA, revealing for the first time, the spectral features of the dust-scattered light which is dominated by late O and early B stars, notably the deep absorption feature near 1000 \AA\ due to overlapping stellar N~III and He~II absorption lines, blended with nearby Si~II and Fe~II and Fe~III lines \citep{Smith_bstar, Smith_ostar}. We confirm the linear drop-off of the CUVB as we approach 912 \AA\ and compare the observed dust-scattered light with our models.

Observations of diffuse line emission have been rare in the UV. The first statistically significant detection of emission from the Galactic halo was of C~IV (1548/1551 \AA) from a Shuttle-based experiment \citep{Martin1990_lines}. This was followed by observations of diffuse line emission of a number of different species from the ultraviolet spectrometer (UVS) on \voyager\ \citep{Murthyvoyovi_2001}, the Far Ultraviolet Spectroscopic Explorer (\FUSE) satellite \citep{Dixon2001, Shelton2001, Welsh2002, Shelton2002, Shelton2003, Otte2003, Otte2006, Shelton2007}, and the Spectroscopy of Plasma Evolution from Astrophysical
Radiation ({\it SPEAR}) satellite \citep{Korpela2006, Welsh2007, Jo2019}. We report emission lines from C~III (977 \AA), O~VI (1032/1038 \AA), and C~IV (1548/1551 \AA) at surface brightnesses similar to previous observations and note, for the first time, a strong emission line at 1135 \AA, likely due to the N~I triplet (1134.2/1134.4/1135 /AA).

\section{Observations}

 The New  Horizons Alice spectrograph \citep{Stern2008} is a Rowland Circle spectrograph with spectral coverage from 520 – 1870~\AA. The detector is an intensified Z-stack micro-channel plate (MCP) with a split coating of KBr (520 -- 1180 \AA) and CsI (1250 -- 1870 \AA) to cover the entire spectral range. The MCP was masked around the \lya\ line (1216~\AA) during the coating process to reduce sensitivity; even so, scattered interplanetary \lya\ emission dominates any diffuse spectrum. The main airglow channel (Fig. \ref{fig:alice_slit}) has two parts: a narrow ``Stem" with a field of view (FOV) of $0\adeg1\times  4\adeg0$ and a square ``Box" with a FOV of $2\adeg0 \times 2\adeg0$. The full-width at half-maximum (FWHM) for aperture-filling diffuse sources is 172~\AA\ for the Box and $9 \pm 1.4$~\AA\ for the Stem \citep{Stern2008a}.
 
 We have described the Alice data and the extraction of the CUVB in \citet{Murthy2025_alice}. The primary contributors to the observed signal are instrumental dark counts, due to fast particles from the onboard radioisotope thermoelectric generator (RTG), and internal scattering of the intense heliospheric \lya\ line across the detector. The dark counts were monitored through dedicated observations interleaved with sky observations, and could be readily subtracted from the data. The \lya\ scattering matrix was measured using differential observations of a single location separated by 16 years (2007 and 2023) and 50 AU (6 AU and 56 AU). The strength of the interplanetary \lya\ line dropped by a factor of 3.5 over that interval and the effects of the line could be measured and subtracted from the observations. The remainder represented the CUVB. The raw data are archived at the NASA Planetary Data System (\url{https://pds-smallbodies.astro.umd.edu/data_sb/missions/nh-kem/index.shtml}) and the reduced Box and Stem spectra used in this work may be downloaded from \citet{Murthy2025_alice}.
 
\begin{table*}[htbp]
\centering
\caption{Stem Observations Used}
\label{tab:obslog}
\begin{tabular}{lccccllll}
\hline
No. & Target & \multicolumn{2}{c}{STEM} & \multicolumn{2}{c}{BOX} & OBS. Date & Exp. Time & E(B - V)\\
& & GL & GB & GL & GB & & s & mag\\
\hline
1 & NCOB05 & 116.06 & -87.85 &  84.36 & -86.79 & 2023.66 &  28800 & 0.012\\
  2 & NCOB07 & 209.75 & -84.38 & 184.71 & -83.37 & 2023.66 &  28800 & 0.018\\
  3 & NCOB06 & 324.18 & -81.82 & 336.32 & -84.42 & 2023.70 &  28800 & 0.015\\
  4 & DCAL01 & 276.07 & -81.65 & 260.77 & -83.94 & 2023.64 &  14400 & 0.022\\
  5 & NCOB9 &  66.93 & -82.49 &  81.04 & -80.57 & 2023.66 &  28800 & 0.014\\
  6 & DCAL02 & 264.89 & -77.99 & 260.38 & -80.80 & 2023.70 &  14400 & 0.023\\
  7 & NCOB10 & 137.93 & -79.83 & 135.27 & -77.06 & 2023.66 &  28800 & 0.015\\
  8 & DARK\_SKY\_2021 &  51.93 & -78.98 &  57.38 & -76.40 & 2021.73 &  28800 & 0.013\\
  9 & NCOB11 & 114.03 & -77.89 & 115.88 & -75.09 & 2023.65 &  28800 & 0.017\\
 10 & NCOB04 & 311.98 & -72.34 & 305.36 & -74.57 & 2023.64 &  28800 & 0.012\\
 11 & NCOB03 & 331.63 & -63.94 & 325.99 & -65.55 & 2023.64 &  28800 & 0.014\\
 12 & NCOB02 & 312.02 & -60.70 & 313.49 & -63.51 & 2023.67 &  28800 & 0.016\\
 13 & NCOB01 & 318.88 & -60.12 & 313.30 & -61.03 & 2023.70 &  28800 & 0.017\\
 14 & NCOB8 &  19.55 & -58.38 &  24.99 & -58.13 & 2023.66 &  28800 & 0.014\\
 15 & DCAL06 & 281.09 & -54.25 & 277.65 & -56.38 & 2023.70 &  14400 & 0.041\\
 16 & NCOB14 &  99.44 & -43.73 & 103.05 & -42.59 & 2023.64 &  28800 & 0.073\\
 17 & NCOB15 &  83.04 &  41.76 &  79.25 &  42.35 & 2023.62 &  28800 & 0.025\\
 18 & DCAL03 &  73.40 &  49.32 &  69.05 &  49.90 & 2023.62 &  14400 & 0.020\\
 19 & DCAL05 &  55.74 &  51.58 &  51.12 &  51.52 & 2023.62 &  14400 & 0.033\\
 20 & DCAL08 &  38.66 &  52.46 &  34.12 &  51.79 & 2023.62 &  14400 & 0.061\\
 21 & DCAL07 &  35.77 &  58.13 &  30.54 &  57.45 & 2023.62 &  14400 & 0.048\\
 22 & SHOCK1 &  26.63 &  58.65 &  21.55 &  57.64 & 2023.62 &  28800 & 0.032\\
 23 & NCOB13 & 344.92 &  60.66 & 341.55 &  58.37 & 2023.63 &  28800 & 0.030\\
24 & NCOB12 & 334.62 &  63.38 & 331.73 &  60.86 & 2023.63 &  28800 & 0.025\\
 25 & SHOCK2 &  14.50 &  66.75 &   8.33 &  65.37 & 2023.62 &  28800 & 0.020\\
 \hline
\end{tabular}
\end{table*}

The observational strategy and planning are detailed in \citet{Murthy2025_alice}. The diffuse UV observations were chosen to study the EBL and, as such, were at high Galactic latitudes ($|b| > 40^{\circ}$) with low reddening (\ebv\ $<$ 0.15 mag), except for one location to search for $H_{2}$ fluorescence. We focused on the Box observations in \citet{Murthy2025_alice} because of their much higher S/N. In this paper, we will study the Stem spectra with their higher spectral resolution. These observations are listed in Table \ref{tab:obslog}, ordered by Galactic latitude.

\section{Results}

\subsection{Emission Lines}

\begin{figure*}
    \includegraphics[width=7in]{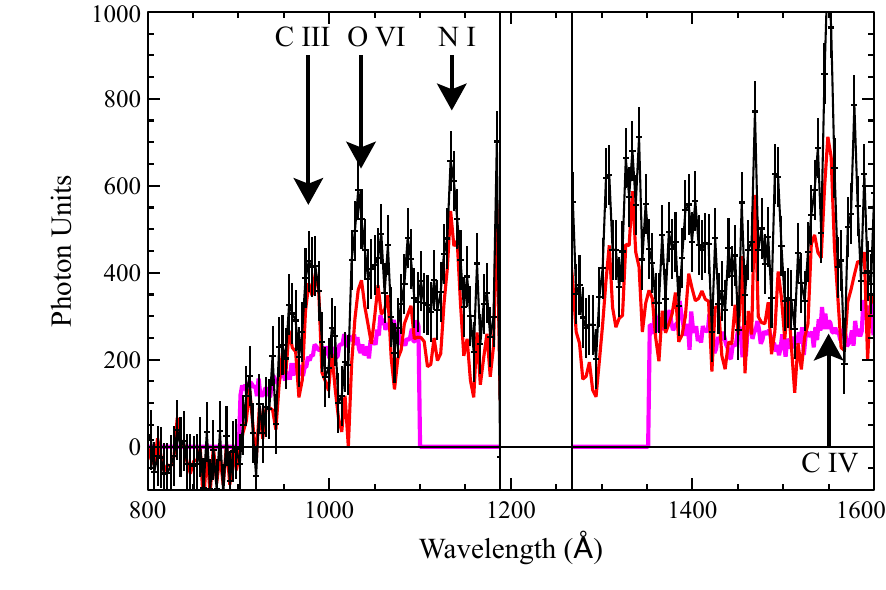}
    \caption{Coadded spectra for the northern (black line with error bars) and southern (red line) Stem observations. The $1 \sigma$ error bar is 69.1 \photu, estimated from the spectral region below 912 \AA. We have blanked out the data around \lya, due to contamination by heliospheric \lya\ emission. The thick magenta line is the offset spectrum as derived from the Box, which has a resolution of about 172 \AA. The positions of the four lines of Table \ref{tab:stem_lines} are shown as arrows.}
    \label{fig:ns_spectra}
\end{figure*}

The signal-to-noise ratio is relatively poor in the individual observations due to the faintness of the diffuse sky, and we have divided the observations into two groups based on latitude. There are 9 observations in the northern hemisphere of the Galaxy ($b > 42^{\circ}$) and 16 in the southern ($b < -42^{\circ}$) with total exposure times of 201,600 and 417,600 seconds, respectively. We further binned the spectra by a factor of 2 in wavelength and plotted them in Fig. \ref{fig:ns_spectra}. We have estimated the $1 \sigma$ error in the data from the standard deviation of the spectra between 600 and 900 \AA, where no astrophysical signal is expected because of interstellar hydrogen absorption. The spectra are essentially identical, except for a difference in continuum level due to the different contributions of the dust-scattered light in each observation. We have also plotted the lower resolution Box spectrum of the offsets from \citet{Murthy2025_alice} for comparison. Note that the relative contribution of the dust-scattered light and the offset will be different for each of the regions.

\begin{figure*}
    \includegraphics[width=6in]{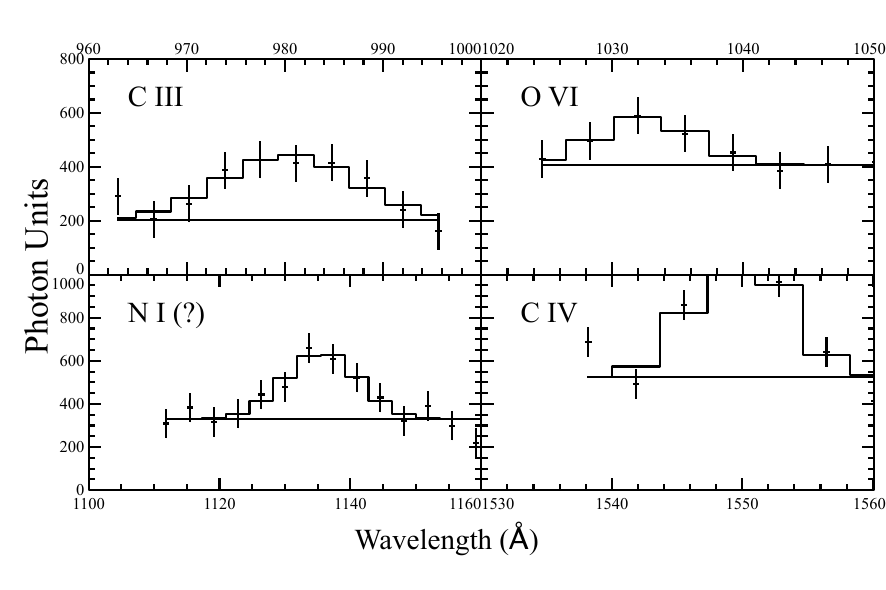}
    \caption{Lines identified in the Stem spectrum from the Northern hemisphere. Data are shown as $1 \sigma$ error bars with the best-fit model and the baseline plotted as solid lines.}
    \label{fig:multilinesn}
\end{figure*}
\begin{figure*}
    \includegraphics[width=6in]{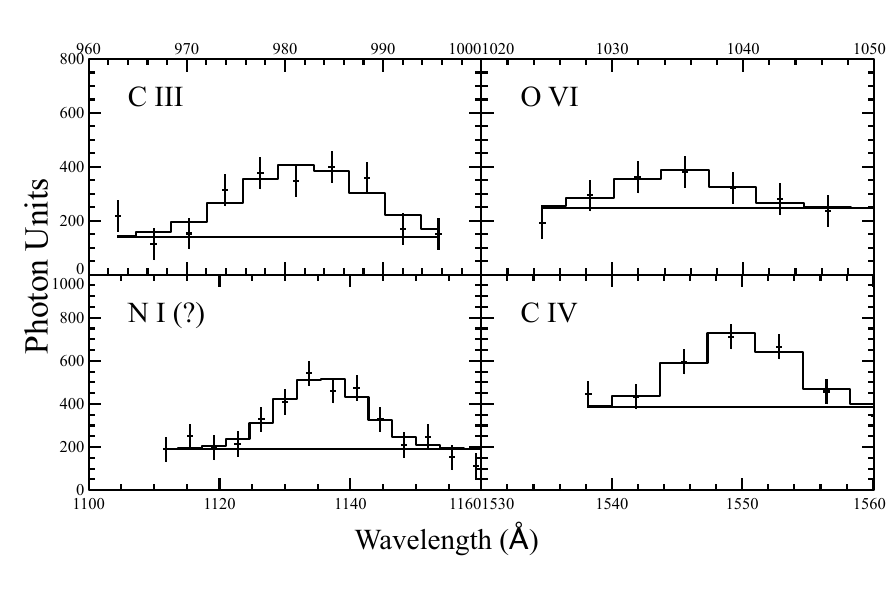}
    \caption{Lines identified in the Stem spectrum from the Southern hemisphere. Data are shown as $1 \sigma$ error bars with the best-fit model and the baseline plotted as solid lines.}
    \label{fig:multiliness}
\end{figure*}

\begin{table}[htbp]
\centering
\caption{Line Strengths}
\label{tab:stem_lines}
\begin{tabular}{lcccc}
\hline
Line & $\lambda_{0}$ & FWHM& \multicolumn{2}{c}{Line Strength }\\
\hline
& \AA & \AA & LU${^a}$ & SI units${^b}$\\
\hline
\multicolumn{5}{c}{North}\\
\hline

CIII &  980.3 &  16.0 & $  4140 \pm 1690 $ & $     8.40 \pm      3.43 $\\
OVI & 1032.4 &   8.6 & $  1630 \pm 1170 $ & $     3.15 \pm      2.26 $\\
NI & 1135.5 &  13.2 & $  4430 \pm 1560 $ & $     7.75 \pm      2.73 $\\
CIV & 1549.8 &   8.3 & $  5520 \pm 1270 $ & $     7.08 \pm      1.63 $\\

\hline
\multicolumn{5}{c}{South}\\
\hline

CIII &  981.9 &  15.3 & $  4320 \pm 1450 $ & $     8.75 \pm      2.93 $\\
OVI & 1034.8 &   9.2 & $  1370 \pm 1250 $ & $     2.64 \pm      2.40 $\\
NI & 1135.7 &  15.4 & $  5460 \pm 1630 $ & $     9.56 \pm      2.86 $\\
CIV & 1549.7 &   9.5 & $  3500 \pm 1210 $ & $     4.48 \pm      1.56 $\\

\hline
\multicolumn{5}{c}{Total}\\
\hline

CIII &  981.4 &  15.4 & $  4200 \pm 1500 $ & $     8.49 \pm      2.98 $\\
OVI & 1033.9 &   8.8 & $  1400 \pm 1300 $ & $     2.69 \pm      2.42 $\\
NI & 1135.6 &  14.8 & $  5150 \pm 1690 $ & $     9.02 \pm      2.95 $\\
CIV & 1549.8 &   8.9 & $  4100 \pm 1200 $ & $     5.24 \pm      1.56 $\\
\hline
\multicolumn{5}{l}{$^a$Line units (photons cm$^{-2}$ s$^{-1}$ sr$^{-1}$)}\\
\multicolumn{5}{l}{$^b$ ($\times 10^{-8}$ ergs cm$^{-2}$ s$^{-1}$ sr$^{-1}$)}\\
\end{tabular}
\end{table}

As discussed above, emission lines have been detected earlier from C~III (977 \AA), O~VI (1032/1038 \AA) and C~IV (1549/1551 \AA) and we searched for emission in the vicinity of those lines, plotted in Fig. \ref{fig:multilinesn} for the Northern hemisphere and in Fig. \ref{fig:multiliness} for the Southern hemisphere. We fit each line with a Gaussian on top of a flat continuum and have tabulated the line parameters and strengths in Table \ref{tab:stem_lines}. Note that, in each case, the continuum has been subtracted before calculating line strengths. Although the FWHM of the Stem is about 9 \AA\, many of the lines are unresolved doublets and we allowed the FWHM in the fits to vary. We will discuss each line below as well as a strong line observable at 1135 \AA, which we have identified with the N~I triplet at 1135 \AA. We did, despite the poor signal-to-noise, search for variability across the individual observations, finding no variation in the line strengths, beyond statistical variations.

\begin{enumerate}
    \item C~III (977 \AA): There are no firm detections of the 977 \AA\ line from the Galactic halo in the literature. \citet{Shelton2002} has placed a $3 \sigma$ upper limit of 3200 LU\footnote{Line Units: ph cm$^{-2}$ s$^{-1}$ ster$^{-1}$}  ; \citet{Shelton2003} observed a $3\sigma$ detection of $4700 \pm 1300$ LU from FUSE observations; and  \citet{Welsh2002} observed a level of $5200 \pm 2000$ LU. Our $\sim 3 \sigma$ detection of $4200 \pm 1500$ LU is consistent with the earlier observations.
    
    \item O~VI (1032/1038 \AA): Typical values of O~VI doublet emission from the Galactic halo are around 3000 -- 4000 LU \citep{Shelton2001, Shelton2002, Shelton2003, Shelton2007, Dixon2001, Dixon2006, Otte2003, Otte2006, Welsh2007, Murthyvoyovi_2001, Jo2019, Korpela2006}. We obtain a marginal detection of $1400 \pm 1300$ LU. This is lower than expected from models of line emission in the Galaxy (see discussion below).
    
    \item C~IV (1548/1551 \AA): Reported measurements of the C~IV doublet range between 4000 -- 6000 LU \citep{Martin1990_lines, Korpela2006, Jo2019}. We detect the C~IV doublet at a $3 \sigma$ level with a value of $4100 \pm 1200$ LU.
    
    \item N~I (1134.2/1134.4/1135.0 \AA): This feature is at the location of the N~I resonance triplet and was seen in both FUSE \citep{Dixon2001, Shelton2001, Shelton2003} and SPEAR spectra \citep{Korpela2006}. However, because N~I emission is not expected from the Galactic halo, this line was ignored in those papers or was identified as either an artifact or an airglow line. We see emission at $3 \sigma$ in all our spectra and, given that we observe the line in all our observations at 56 AU, it must be real and from Galaxy emission.

    \item There may be other lines visible, such as the O~I lines at 1304 and 1356 \AA\ or the C~II line at 1334 \AA; however, the data are noisy and their existence will have to be confirmed through observations from other instruments such as FUSE or the Cosmic Origins Spectrograph (COS) on the {\it Hubble Space Telescope}.
    
\end{enumerate}

\subsubsection{O~VI discussion:}

Our quoted measurement of O~VI emission was derived from combining multiple fields at high Galactic latitude ($|b| > 42^{\circ}$), with 9 fields in the northern hemisphere and 16 fields in the southern hemisphere.  A survey with the FUSE spectrograph \citep{Dixon2006} detected diffuse emission from the O~VI doublet (1031.926\AA\ and 1037.617~\AA) in multiple locations.  Out of 183 total sight lines, 29 yielded $>3\sigma$ detections with a median surface brightness of $I_{\rm OVI} =  3300$~LU.  An additional 35 sight lines provided upper limits of 2000 LU or less.  They also noted that  21 O~VI-emitting regions at high Galactic latitude exhibited lower intensity than those at low latitudes.  
Using the methodology of \citet{Shull1994}, they combined the O~VI emission
measurements with published O~VI absorption column densities to estimate the electron density $n_e$ in a cloud-hot-gas interface in the ISM. They derived 
$n_e \approx$ 0.2--0.3~cm$^{-3}$ and an interface length $L \approx 0.1$~pc.  

Although the surface brightness of O~VI in the stacked Alice data is below that seen in many of the FUSE sight lines, it may reflect low electron densities and high plasma temperatures in the high-latitude fields observed by New Horizons.  For a planar slab at distance $d$, with projected area $A$ and length $L$,  the O~VI surface brightness averaged over solid angle $(A/d^2)$ can be written as
\begin{equation}
  I_{\rm OVI} = \frac {n_e \, n_{\rm OVI} \langle \sigma v \rangle (AL)} {4 \pi d^2 (A/d^2) }
     = \frac {n_e n_{\rm OVI} \langle \sigma v \rangle L  }{4 \pi} \; .  
 \end{equation} 
The Maxwellian-averaged electron-impact excitation rate coefficient of the  O~VI doublet is 
\begin{equation}
  \langle \sigma v \rangle = (8.629\times10^{-6}~{\rm cm}^3~{\rm s}^{-1})
      \frac {\Omega_{12}}{g_1 T^{1/2}}  \exp (-E_{12}/kT) \; .
 \end{equation}
 We assume a collision strength $\Omega_{12} \approx 5$ at $T \approx 10^5$~K
 \citep{Shull1994} and mean excitation energy $E_{12} = 12.01$~eV of the 
 1034.82~\AA\ doublet (with $g_1 = 2$) to find 
 $\langle \sigma v \rangle \approx 1.7\times10^{-8}{\rm cm}^3~{\rm s}^{-1}$.
 For hot, fully ionized interstellar gas with solar oxygen abundance 
 O/H $\approx 5.62\times10^{-4}$ \citep{bergemann2021} and $n_e = 1.2 n_{\rm H}$, 
 we estimate an O~VI surface brightness of
  \begin{equation}
    I_{\rm OVI} \approx (2440~{\rm LU}) \left( \frac {n_e}{0.25~{\rm cm}^{-3}} \right)^2 
         \left( \frac {L}{0.1~{\rm pc}} \right) \left( \frac {f_{\rm OVI}} {0.2} \right) \; ,
  \end{equation}  
 where we scaled to OVI\ ionization fraction $f_{\rm OVI} = 0.2$, interface 
 length $L \approx 0.1$~pc, and $n_e = 0.25~{\rm cm}^{-3}$.  
 
 Thus, the undetected regions with $I_{\rm OVI} < 2000$~LU could represent 
 cloud interfaces with low density $n_e < 0.1~{\rm cm}^{-3}$ or regions in which 
 $f_{\rm OVI} < 0.1$.  For example, little O~VI emission is expected from the interior  of the local hot bubble \citep{yeung2024}, owing to its very low density 
 $n_e \approx 0.004~{\rm cm}^{-3}$ and high temperature $T \approx 10^6$~K.  
 At these temperatures, O~VI has a small ionization fraction, 
 $f_{\rm OVI} \approx 3.5 \times10^{-3}$.  Even over a large bubble pathlength 
 $L \approx 85$~pc, we estimate that $I_{\rm OVI} \approx 10$~LU.  

\subsection{Dust-Scattered Light}

\subsubsection{Modeling}

Our modeling procedure is based on \citet{Murthy2025}. The CUVB is given by:
\begin{equation}
    CUVB = D(a, g) + O \times e^{-\tau}
    \label{eq:cub_eq}
\end{equation}
where:
\begin{itemize}
    \item CUVB is the total observed diffuse signal (\photu).
    \item D is the dust-scattered light from our model as a function of the optical constants ($a$ and $g$) and the \ebv\ in the line of sight, to be described further below.
    \item O is the offset, assumed to be uniform and extincted by the total reddening in the line of sight. It is comprised of the EBL and contributors from the Galactic halo.
\end{itemize}

The first step in calculating the dust-scattered light (D) is to calculate the contribution of the individual stars at the location of scattering. The stars were taken from the Hipparcos catalog \citep{Perryman1997}, which includes their locations and spectral types. The contribution of each star was modeled using TLUSTY spectral profiles \citep{Lanz2003, Lanz2007}. We used the 3-dimensional dust map of \citet{Green2019} to calculate the dust between each star and the point of interest and extincted the starlight using the extinction curve of \citet{Draine_scat2003}.

The dust distribution in the line-of-sight was assumed to fall off exponentially with a scale height of 125 pc \citep{Marshall2006} and a cavity of radius 50 pc around the Sun \citep{Welsh2010} and we used the Planck \ebv\ \citep{PlanckDust2016} for the total column density. We divided the line-of-sight into 5 pc bins and calculated the contribution at the Earth from each star for each bin for a given set of optical constants. The total dust-scattered light is the sum over all stars and all bins. Because most of the light is from the back-scattering of stars in the Galactic Plane, the scattered light is much higher when the dust-scattering is closer to isotropic (smaller $g$).

\begin{figure}
    \includegraphics[width=3in]{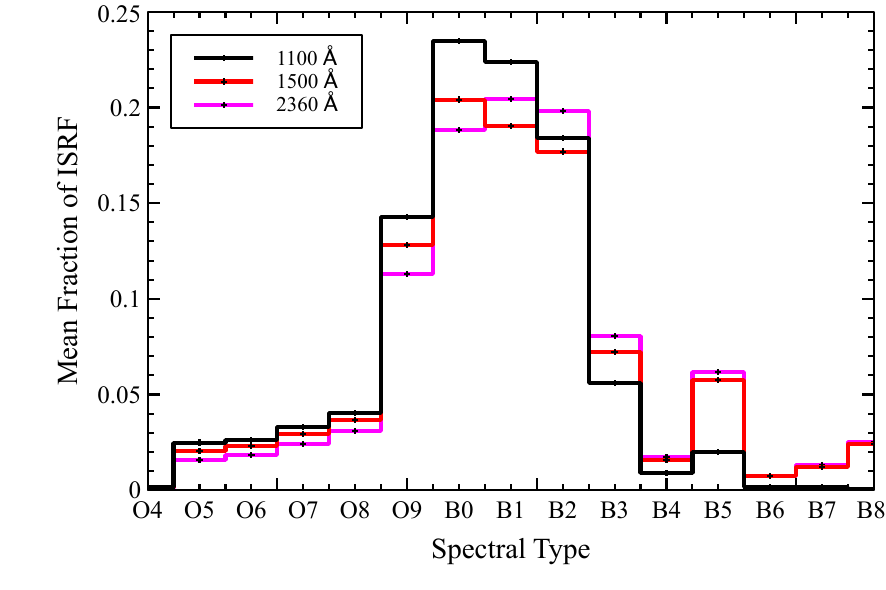}
    \caption{Contribution to scattered light at 1100 \AA\ (black line), 1500 \AA\ (red line), and 2360 \AA\ (magenta line) from stars of different spectral types, averaged over all observations.}
    \label{fig:star_contrib}
\end{figure}

\begin{figure}
    \includegraphics[width=3in]{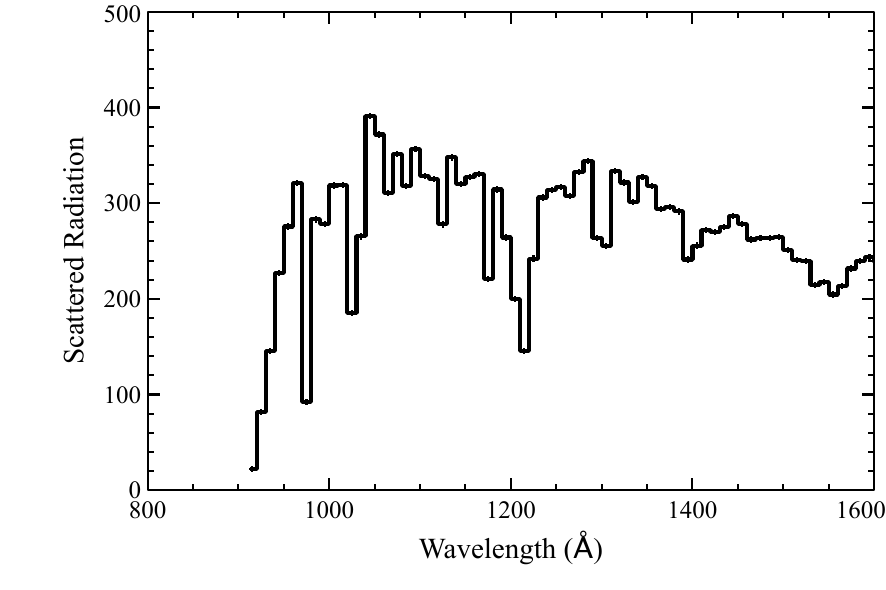}
    \caption{Representative dust-scattered spectrum in photon units. The scaling is arbitrary.}
    \label{fig:isrf_spectra}
\end{figure}

 The ISRF in each of the locations is dominated by a few stars ($< 100$) with spectral types from O9 to B2 (Fig. \ref{fig:star_contrib}) and this is reflected in the typical model spectrum of the ISRF (Fig. \ref{fig:isrf_spectra}). The deep absorption feature near 1000 \AA\ is due to overlapping stellar N~III and He~II absorption lines, blended with nearby Si~II and Fe lines \citep{Smith_bstar, Smith_ostar}, and is visible in the observed spectrum (Fig. \ref{fig:ns_spectra}).

\subsubsection{Application to the data}

\begin{figure}
    \includegraphics[width=3in]{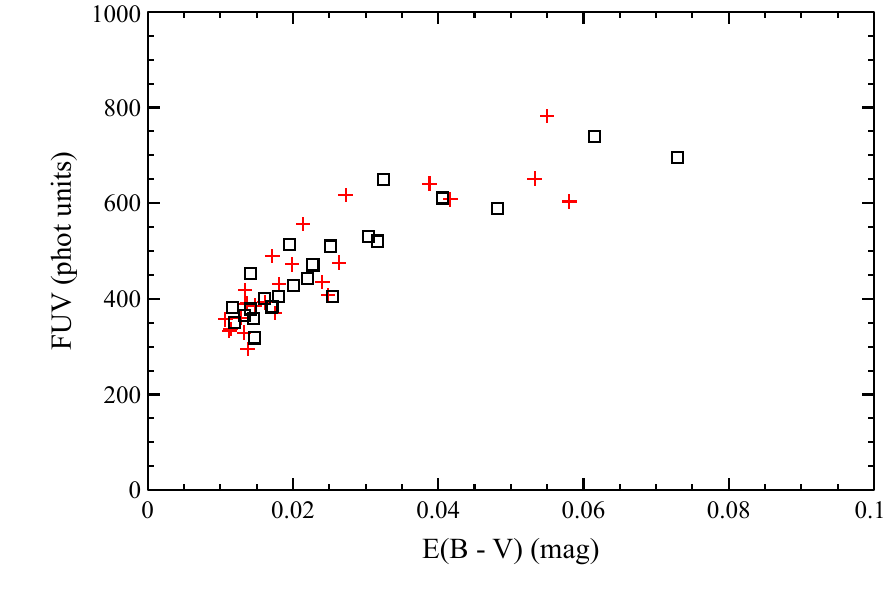}
    \includegraphics[width=3in]{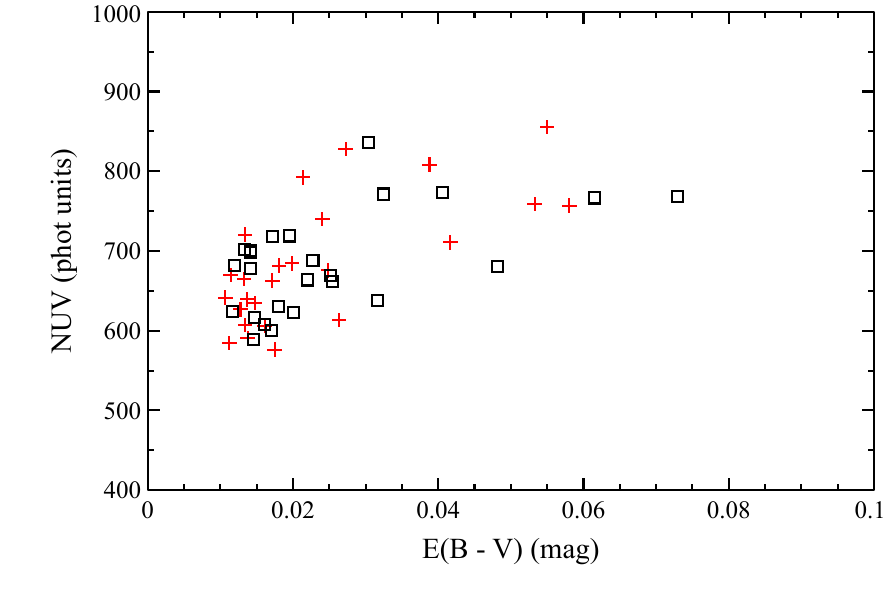}
    \caption{\galex\ FUV (top) and NUV (bottom) surface brightness as a function of \ebv. Values from the Stem locations are shown as black squares; values from the Box locations are shown as red crosses. The correlation coefficients are r = 0.887 and r = 0.619 for the FUV-\ebv\ and NUV-\ebv, respectively.}
    \label{fig:ebv_galex}
\end{figure}

We first applied our models to \galex\ FUV and NUV data at the NH target locations. The CUVB observations were from \citet{Murthy2014apj} with \ebv\ from Planck \citep{PlanckDust2016}, and are averaged over the Stem and Box, respectively (Fig. \ref{fig:ebv_galex}). 

\begin{figure}
    \includegraphics[width=3in]{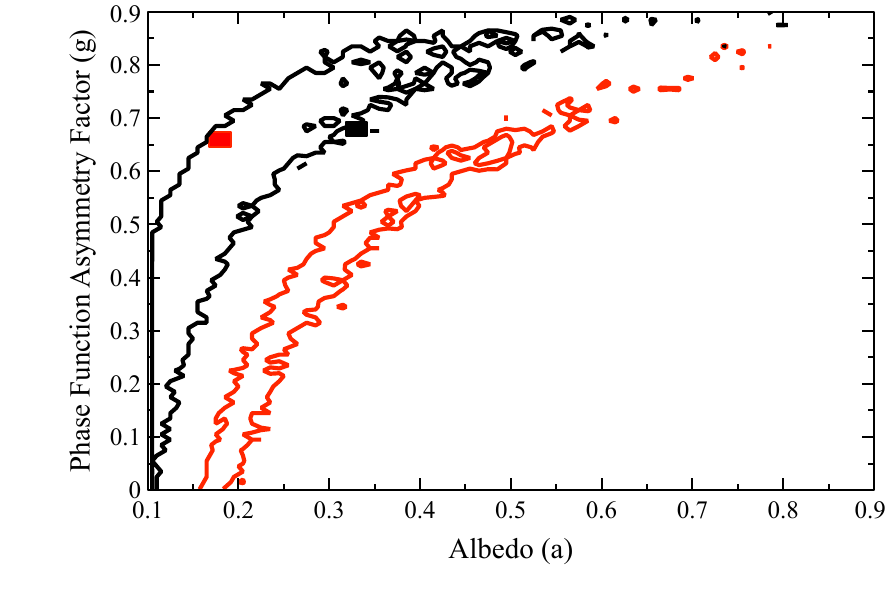}
    \caption{The allowed optical constants at the $1 \sigma$ level in the FUV (black contours) and NUV (red contours) from the \galex\ observations. The predicted values from \citet{Draine_scat2003} are shown in the FUV (black square: $a =0.33, g = 0.68$) and the NUV (red square: $a = 0.18, g = 0.66$). Note that we have not allowed the albedo to fall below 0.1.}
    \label{fig:ag}
\end{figure}

\begin{figure*}
    \includegraphics[width=6in]{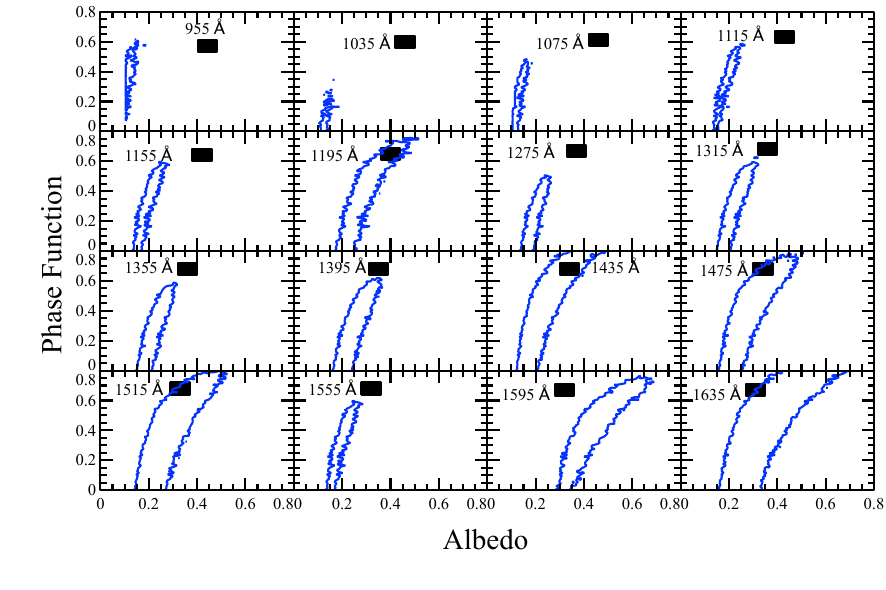}
    \caption{$1 \sigma$ contours for the optical constants from the Stem spectra. The data have been binned by 40 \AA\ and exclude the region around \lya. The predicted values from the astronomical silicate model of \citet{Draine_scat2003} are plotted as black squares.}
    \label{fig:stem_ag}
\end{figure*}

We fit Eq. \ref{eq:cub_eq} to the \galex\ FUV and NUV data and placed $1\sigma$ limits on the optical constants (Fig. \ref{fig:ag}) using the formulation of \citet{Lampton1976}. We extended this analysis to the Stem spectra, which we had binned by 40 \AA\ to increase the signal-to-noise and have plotted the resultant $a-g$ contour plots in Fig. \ref{fig:stem_ag}. The optical constants are broadly consistent with the predicted values of \citet{Draine_scat2003} with differences at shorter wavelengths. It is, however, important to note that the results may not be physically meaningful as the Henyey-Greenstein \citep{Henyey1941} function, itself, has no physical basis. Furthermore, as \citet{Mathis2002} noted, models of the dust scattering are not unique and can trade albedo for $g$; i.e., models with high albedo and highly forward-scattering grains are identical, within the limits of the data, to models with low albedo and isotropically scattering grains.

\subsubsection{Offsets}

\begin{figure*}
    \includegraphics[width=7in]{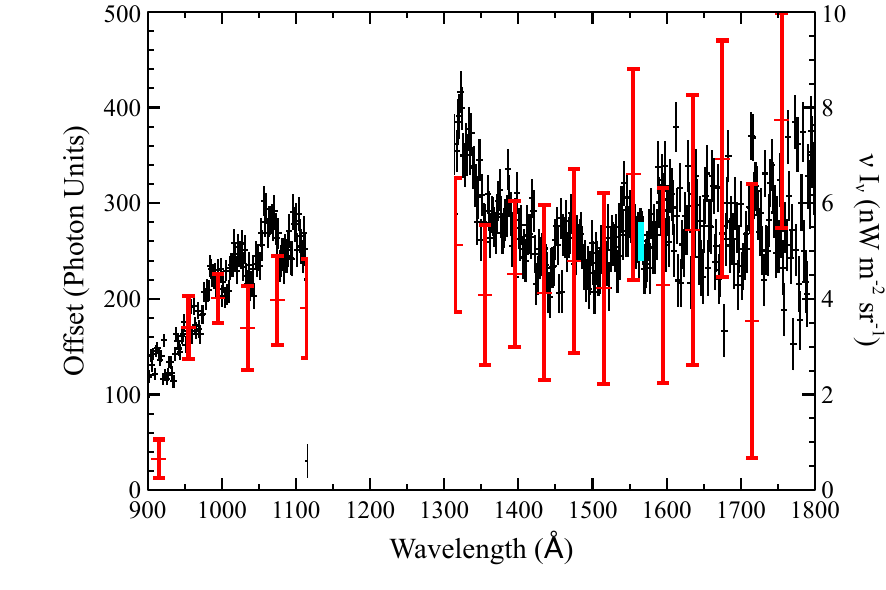}
    \caption{$1 \sigma$ error bars (red bars) from the Stem observations binned to a 40 \AA\ in wavelength. The offsets from the Box observations \citep{Murthy2025_alice} are plotted as $1 \sigma$ black error bars with  the \galex\ limit of \citet{Akshaya2019} plotted as the blue bar. This spectrum may be compared with Fig. \ref{fig:ns_spectra} to see how much of the total CUB near the Poles is from the offset radiation.}
    \label{fig:stem_offset}
\end{figure*}

We had determined the spectral shape of the offsets using the Box spectra in \citet{Murthy2025_alice}, which indicated a linear decline toward shorter wavelengths. This interpretation, however, was potentially compromised by the Box's broad spectral resolution (170 \AA). The convolution of the signal longward of 912 \AA\ with the zero emission shortward of the Lyman limit would necessarily have led to the observed behavior. The Stem data offer an independent determination of the offsets as a function of wavelength at a much better spectral resolution. We have rebinned the Stem spectra to 40 \AA\ to increase the signal-to-noise and have plotted the resultant $1\sigma$ offsets in Fig.~\ref{fig:stem_offset}. We find that the Stem results are consistent with the Box results, albeit with much larger uncertainties, showing that there is indeed a drop-off in the offsets at shorter wavelengths, although still non-zero.

\section{Conclusions}

We have used observations from the Stem aperture of the New Horizons Alice spectrograph to study the cosmic ultraviolet background at a spectral resolution of 9 \AA. We observed C~III (977 \AA) and C~IV (1548/1551 \AA) at $3\sigma$ confidence limits of $4200 \pm\ 1500$ and $4100 \pm 1200$ LU, respectively, and obtained a marginal detection of O~VI (1032/1038 \AA) at $1400 \pm 1300$ LU. These are consistent with earlier detections from \voyager, FUSE, and SPEAR. We observed a strong line at 1135 \AA\, which we have identified with the N~I triplet emission. This line was observed earlier but had been speculated to be due to airglow or instrumental effects.

The dust-scattered light is primarily from a small number of hot stars whose spectral lines are seen in the scattered spectrum. We find that the albedo of the scattering dust grains is low ($ a < 0.6)$ throughout the entire spectral region from 900 -- 1600 \AA. The results are broadly consistent with the astronomical silicate model of \citet{Draine_scat2003}. The offsets confirm the spectral behavior seen in the Box, with a dropoff to the Lyman limit. This result reinforces the observational case for a significant new component to the diffuse UV background, one whose origin remains as yet unexplained \citep{Henry2015}. It cannot be attributed to the annihilation or decay of traditional dark-matter candidates including WIMPs, axions or primordial black holes \citep{Akshaya2018}. It could, however, be evidence for Axion Quark Nuggets (AQNs), newer candidates whose existence would not only provide the needed cold dark matter but also address the coincidence problem (similar densities of dark and baryonic matter) and the origin of baryon asymmetry \citep{Zhitnitsky2022}. AQNs contribute to the diffuse Galactic UV background via thermal Bremsstrahlung radiation when they scatter off protons in the interstellar medium \citep{Sekatchev2026, vanwaerbeke2026}. The predicted signal has a broad spectrum (like the offset which we report here) and tracks primarily with the density of ionized gas; not dust, consistent with the observed lack of correlation with the Galactic infrared background or distribution of hot UV-emitting stars.

The vantage point of New Horizons near the edge of the Solar System, beyond much of the heliospheric \lya\ emission, has allowed new determinations of the diffuse background in the optical \citep{Symons2023, Postman2024} and in the UV \citep{Murthy2025_alice, Gladstone2025}. These are not possible except from locations beyond Pluto \citep{zemcov_nh2018} and have provided valuable new insights on the CUVB.

\section*{Acknowledgments}
We thank NASA for funding and support of the New Horizons mission. The data presented were obtained during the second Kuiper Extended Mission of New Horizons. Some of the data presented in this paper were obtained from the Mikulski Archive for Space Telescopes (MAST). STScI is operated by the Association of Universities for Research in Astronomy, Inc., under NASA contract NAS5-26555. Support for MAST for non-HST data is provided by the NASA Office of Space Science via grant NNX13AC07G and by other grants and contracts. This research has made use of the SIMBAD database, CDS, Strasbourg Astronomical Observatory, France. We thank the referee, Stephan McCandliss, for comments and discussion which greatly improved and clarified the concepts in the paper.
\vspace{-1em}


\bibliography{murthy}{}
\bibliographystyle{mnras}

\end{document}